\documentclass[aps,prb,reprint,groupedaddress]{revtex4-2}

\usepackage{todo} 
\usepackage{graphicx} 
\usepackage{dcolumn}  
\usepackage{amsmath}  
\usepackage{amssymb}  
\usepackage{color} 
\usepackage{hyperref} 

\graphicspath{{./figures/}}

\begin{document}

\title{Quasicrystalline chiral soliton lattices in a Fibonacci helimagnet}
\author{Pablo S. Cornaglia }
\email[]{pablo.cornaglia@cab.cnea.gov.ar}
\affiliation{Centro Atómico Bariloche and Instituto Balseiro, CNEA, 8400 Bariloche, Argentina}
\affiliation{Consejo Nacional de Investigaciones Científicas y Técnicas (CONICET), Argentina}
\affiliation{Instituto de Nanociencia y Nanotecnología CNEA-CONICET, Argentina}

\author{Leandro M. Chinellato}
\affiliation{Department of Physics and Astronomy, The University of Tennessee, Knoxville, TN, 37996, USA}

\author{Cristian D. Batista}
\affiliation{Department of Physics and Astronomy, The University of Tennessee, Knoxville, TN, 37996, USA}
\affiliation{Neutron Scattering Division and Shull-Wollan Center, Oak Ridge National Laboratory, Oak Ridge, TN, 37831, USA}

\date{\today}

\begin{abstract}
We investigate the ground state magnetic configurations of a Fibonacci chain of classical spins with nearest-neighbor ferromagnetic and monoaxial Dzyaloshinskii-Moriya exchange interactions. Our analysis reveals a diverse array of magnetic textures induced by an external magnetic field perpendicular to the Dzyaloshinskii-Moriya vector. 
These textures exhibit a spectrum ranging from a quasi-fully polarized non-collinear state under high magnetic fields, capable of maintaining metastable chiral soliton topological defects, to a variety of quasicrystalline chiral soliton lattices below a critical field $H_c$.
For a range of magnetic fields below $H_c$, the ground state spin textures result from the interplay between an effective quasiperiodic potential influencing the solitons and their repulsive interactions. At lower magnetic fields, the system experiences a commensurate-incommensurate transition, signified by the appearance of discommensurations in the quasicrystalline soliton lattice. In the absence of an external magnetic field, the ground state assumes a helical configuration with a quasiperiodic pitch angle.
\end{abstract}

\maketitle

\section{Introduction}
Non-collinear magnetic phases arise as a consequence of frustration effects, driven either by the lattice geometry or competing interactions. These magnetic textures present a rich variety of physical phenomena including topological defects, such  as skyrmions and chiral solitons~\cite{kishine2015theory,Nishikawa2016,ghimire2013magnetic,laliena2023continuum,Osorio2022}, topological magnon bands \cite{mcclarty2022topological}, anomalous Hall effect in metals \cite{nakatsuji2015large} and enhanced magnetoelectric effects in insulators~\cite{brataas2006non}. 
The sensitivity of these systems to external magnetic fields and currents facilitates the manipulation and control of the spin textures and their magnon spectrum, enabling potential applications in data transmission and storage for spintronic devices~\cite{brataas2006non,chumak2015magnon}.

Quasicrystalline structures offer a promising avenue to explore new facets of frustration and enable the observation of novel non-collinear magnetic phases~\cite{cornaglia2023unveiling}. Characterized by their aperiodic structure, quasicrystals exhibit distinctive Bragg diffraction peaks without possessing translational symmetry. Magnetism in quasicrystals has been investigated as a method to tailor both the magnon spectrum~\cite{lisiecki2019magnons, costa2013band, coelho2010quasiperiodic, gubbiotti2016collective, rychly2015spin, rychly2016spin, gubbiotti2018reprogrammable} and  lifetime~\cite{jeon2022fractalized}. Despite these efforts, the field, especially the study of topological defects in magnetic quasicrystals, still remains largely uncharted.

Chiral solitons in helimagnets have garnered significant attention in recent years~\cite{kishine2015theory, Osorio2022, Nishikawa2016, ghimire2013magnetic}. These solitons are topologically protected, rendering them stable under perturbations. The observation of the chiral soliton lattice in the hexagonal helimagnet CrNb$_3$S$_6$ marked a fundamental milestone~\cite{togawa2012chiral}, more than 50 years after its theoretical prediction~\cite{dzyaloshinskii1964theory,de1968calcul}. This material features ferromagnetic layers coupled by exchange and Dzyaloshinskii-Moriya interactions, and helimagnetism arises from the competition between these two interactions. Under an external magnetic field ${\vec{H}}$ perpendicular to the helimagnet axis, a soliton lattice emerges, with a lattice parameter that can be tuned by varying the magnetic field intensity.

To investigate the interplay between quasicrystalline structures and magnetism, with a particular focus on topological defects, we employ a chiral helimagnetism model on a Fibonacci quasicrystal. The Fibonacci quasicrystal stands out as one of the extensively studied quasiperiodic structures, thoroughly characterized in existing literature~\cite{baake2023fibonacci, baake2013aperiodic, Jagannathan21, huang2015sequence}. Our findings unveil a diverse array of magnetic textures, including various quasicrystalline chiral soliton lattices. Additionally, we observe field induced transitions between these lattices, as well as a commensurate-incommensurate transition.

The remainder of the paper is organized as follows. In Section \ref{sec:model}, we introduce the model for chiral helimagnetism in a Fibonacci quasicrystal and provide a description of known results for a regular chain. Section \ref{sec:results} begins with the characterization of the zero and high field regimes, followed by an exploration of single chiral soliton states. We then analyze the structure of various chiral soliton lattices that emerge in response to the external magnetic field. In Section \ref{sec:phasediagram}, we present the phase diagram as a function of the magnetic field. Finally, in Section \ref{sec:conclusions} we  summarize our main results and conclusions.

\section{Model and Methods} \label{sec:model}

We consider an effective one dimensional classical spin Hamiltonian for chiral helimagnetism\cite{kishine2015theory} 
\begin{equation}\label{eq:totenergy}
H = \sum_{i=1}^{N-1} J_{i}\,\vec{S}_i \cdot \vec{S}_{i+1}+ \sum_{i=1}^{N-1} \vec{D}_{i}\cdot(\vec{S}_i \times \vec{S}_{i+1})- \vec{H}\cdot \sum_{i=1}^{N} \vec{S}_i,
\end{equation}
where the spin $\vec{S}_i$  at site $i$ is  represented by a unit axial vector. The interaction terms include a ferromagnetic exchange interaction $J_{i}<0$, a Dzyaloshinskii-Moriya (DM) interaction with a DM vector $\vec{D}_{i}=D_{i}\hat{x}$ parallel to the $\hat{x}$ axis. 
The last term represents a Zeeman coupling to the  external magnetic field $\vec{B}= \vec{H}/ g \mu_B$, where  $\vec{H}=H\hat{z}$ is perpendicular to the DM vector, $g$ is the Land\'e factor and $\mu_B$ is the Bohr magneton.

\subsection{Homogeneous chain}

In the absence of an external magnetic field, the  ground state of the homogeneous system ($D_{i}=D$ and $J_{i}=J$) is a chiral helix.  Since the DM interaction favors a polarization plane  perpendicular to $\hat{x}$:
$\vec{S}_i=(0,\sin \phi_i,\cos \phi_i)$. The chirality of the helix is determined by the sign of $D$  and the pitch angle $\phi_{i+1}-\phi_i=\arctan\left(D/J\right)$ determines the  characteristic wavevector of the helix.

For small pitch angles, $|\phi_i - \phi_{i+1}| \ll 1$, we can take the long wavelength (continuum) limit of the lattice model. The resulting effective Hamiltonian is a  sine-Gordon model, whose
analytical ground state solutions  are  chiral soliton lattices with a spatial period that approaches infinity for  $ H \to H_c$. When the magnetic field exceeds $H_c$ ($H > H_c$), the system transitions to a fully polarized phase~\cite{dzyaloshinskii1964theory,de1968calcul}.
Back to the lattice, the single-soliton solution can be written as
\begin{equation}
    \phi_i = 4 \arctan(e^{(i-x_s)/l_s}),
\end{equation}
where $x_s$ is the soliton position in units of the lattice parameter and $l_s(H)$ is a field dependent characteristic length.
The solitons can be interpreted as extended particles with a repulsive two-body interaction that diminishes exponentially with increasing distance.

Due to the effective easy-plane anisotropy, the projection of the spin configuration onto the $yz$ plane establishes a mapping $f: S^1 \to S^1$ in the continuum limit. The count of solitons is determined by the topological degree of this mapping, which corresponds to the winding number.
In the context of long-wavelength structures (small pitch angle), we can calculate the winding number of the spin configuration on the lattice by employing the geodesic interpolation:
\begin{equation}
N_s= \frac{1}{2\pi}\left|\sum_{i=1}^{N-1} \arcsin\left[\hat{x}\cdot(\vec{S}_i\times \vec{S}_{i+1})\right]\right|.
\end{equation}
\subsection{Fibonacci chain}

In the subsequent discussion, we focus on spins within a Fibonacci chain. This chain comprises two types of bonds, denoted as $S$ and $L$, arranged in a quasiperiodic fashion. The construction of the Fibonacci chain follows a straightforward inductive concatenation rule. The sequence initiates with single-bond chains $\Omega_1=S$ and $\Omega_2=L$. Each subsequent chain $\Omega_m$ is formed by concatenating the two preceding ones, 
\begin{equation}\label{eq:concat}
\Omega_m=\Omega_{m-1}\oplus \Omega_{m-2},
\end{equation} 
as indicated in  Table \ref{tab:fibo}.
\begin{table}[t] 
\centering
\caption{Finite Fibonacci chains}
\label{tab:fibo}
\begin{tabular}{c|l|c} 
$m$&$\Omega_m$=$\Omega_{m-1}\oplus \Omega_{m-2}$&$F_m=F_{m-1}+F_{m-2}$\\
\hline
\hline
1&$S$&1\\
2&$L$&1\\
3&$LS$&2\\
4&$LSL$&3\\
5&$LSLLS$&5\\
6&$LSLLSLSL$&8\\
7&$\underbrace{LSLLSLSL}_{\Omega_6}\underbrace{LSLLS}_{\Omega_5}$&13\\
\end{tabular}
\end{table}
The number of bonds in $\Omega_m$ is the Fibonacci number $F_{m}$ and the corresponding number of spins is $N=F_m+1$.  The ratio of two consecutive Fibonacci numbers converges to the golden ratio $\tau=(1+\sqrt{5}) / 2$ for large $m$: 
$$\lim_{m\to\infty}F_m/F_{m-1}\to\tau,$$
and so does the ratio of the number of $L$ and $S$ bonds in the chain.

The Fibonacci chain can also be constructed applying repeatedly the inflation rule 
\begin{align}
S &\to L\nonumber \\
L &\to LS\\
\end{align}
to a starting $S$ bond.
The exchange interaction  $J_{i}$ and the DM vector $\vec{D}_{i}$ in Eq.~\eqref{eq:totenergy} assume values based on the type of bond connecting spins $\vec{S}_i$ and $\vec{S}_{i+1}$. Specifically, $J_{i}$ takes the value $J_L$ for an $L$ bond and $J_S$ for an $S$ bond, while $\vec{D}_{i}$ analogously corresponds to either $D_L \hat{x}$ or $D_S \hat{x}$, meaning both coupling constants follow a quasiperiodic pattern. 
In what follow we take $|J_S|=1$ as the energy unit.

\section{Ground states of the Fibonacci chain} \label{sec:results}

In the absence of the external magnetic field, the spin Hamiltonian \eqref{eq:totenergy} is invariant under global spin rotations along the ($\hat{x}$) direction of the DM vector.
In other words, the energy depends only on the relative angle between consecutive spins along the chain.  
The ground state can then be constructed as follows: 
\begin{enumerate}
\item  Initializing the first spin on the chain to an arbitrary direction perpendicular to the DM vector. 
\item  Setting the orientation of its nearest neighbor in order to minimize the interaction energy between the two spins. This is obtained for a relative angle $\phi_\alpha=\arctan(D_\alpha/J_\alpha)$ for spins joined by an $\alpha=\left\{L,S\right\}$ bond. 
\item  Repeating the second step sequentially for all spins along the chain. 
\end{enumerate}
The resulting spin configuration  is a helix with a polarization plane perpendicular to the $\hat{x}$-axis~\footnote[1]{The sign of $D_{L,S}$  determines the chirality of the helix.} and the spin  orientations  follow the quasiperiodic pattern of the Fibonacci chain. The angle of the $i$-th spin is obtained counting the number of $S$ bonds $N_S(i)$ and of $L$ bonds $N_L(i)$ up to position $i=1+N_S(i)+N_L(i)$
\begin{align}
\phi_i &=\phi_1 + N_S(i)\phi_S +N_L(i)\phi_L.
\end{align}
 where $\phi_1$ is the angle of the first spin on the chain. The relative angle of two spins $\vec{S}_i$ and $\vec{S}_j$  separated by a Fibonacci number of bonds $j-i=F_n\gg 1$ can be approximated by $\phi_j-\phi_i\simeq F_n \phi_{\rm{avg}}\mod(2 \pi)$, where
 $$\phi_{\rm{avg}}=\frac{(\phi_S+\tau\phi_L)}{1+\tau}$$
 is the average pitch angle of the Fibonacci chain.
This can be shown by using the relations  $N_S(j)-N_S(i)=N_S(i+F_n)-N_S(i)\simeq F_{n-2}$ and $N_L(j)-N_L(i)=N_L(i+F_n)-N_L(i)\simeq F_{n-1}$ which lead to
\begin{align}
 \phi_j-\phi_i &\simeq \left[F_n \left(\frac{F_{n-2}}{F_n}\phi_S +\frac{F_{n-1}}{F_n}\phi_L\right)\right]  \!\!\!\!\!\mod(2\pi)\nonumber\\
&\simeq  \left[F_n \left(\frac{\phi_S}{\tau^2} +\frac{\phi_L}{\tau}\right)\right] \!\!\!\!\! \mod(2\pi)
\\\nonumber&= F_n\phi_{\rm{avg}} \!\!\! \mod (2\pi), \nonumber
\end{align}
where we have used the relation $\tau^2= 1+\tau$.
This state coincides with the spin configuration obtained from numerical minimization of the magnetic energy presented in Fig.~\ref{fig:3dstates_helical}. The details of the numerical calculations are presented in Appendix \ref{app:minim}.
\begin{figure}[t]
\includegraphics[width=\columnwidth]{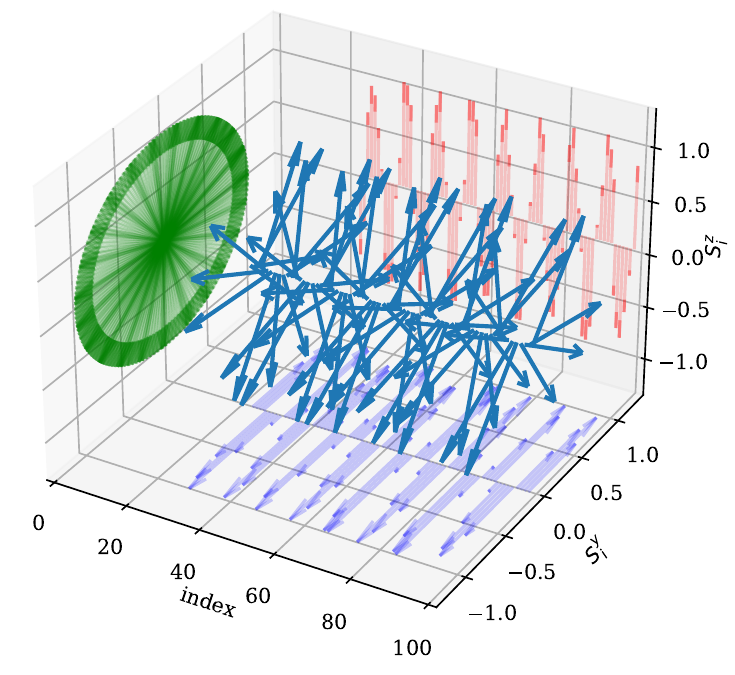}
\caption{Helical ground state configuration at $H=0$, $J_S=-1.0 $, $J_L=-1.75$, $D_S=-2.25$, and $D_L=-0.75$. 
}
\label{fig:3dstates_helical}    
\end{figure}

\subsection{Quasi-fully  polarized phase}

\begin{figure}[t]
\includegraphics[width=\columnwidth]{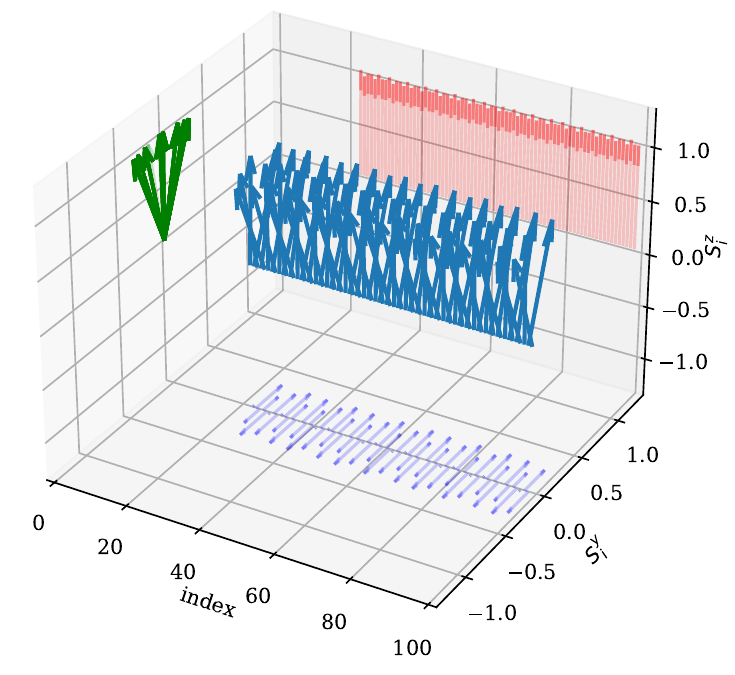}
\caption{Quasi-fully polarized configuration at $H=0.75$. Other parameters as in Fig. \ref{fig:3dstates_helical}. }
\label{fig:3dstates_polarized}    
\end{figure}

For $H>H_c$, the ground state configuration is devoid of solitons ($N_s=0$), analogous to the case in a homogeneous chain.
As illustrated in Fig. \ref{fig:3dstates_polarized}, however,  the quasiperiodic structure of the magnetic interactions, in particular of the DM couplings, makes the ground state of the Fibonacci chain quasi-fully polarized (QFP), non-collinear and quasiperiodic.
To understand the ground state structure of the Fibonacci chain in this regime, it is instructive to initially consider the case of a regular lattice with alternating couplings $D_L$ and $D_S$. It can be demonstrated straightforwardly that the ground state  is non-collinear, and the spin configurations can be parameterized, for large $H$, with angles 
$$\phi_i \sim \frac{(-1)^i(D_L-D_S)}{(H-2J_L-2J_S)}.$$
Despite the DM interaction energy having different signs for $L$ and $S$ bonds, it is not compensated if $D_L\neq D_S$, favoring the generation of a non-collinear state. For homogeneous DM interactions ($D_L=D_S$), the ground state is fully polarized, optimizing both the ferromagnetic exchange and the Zeeman energies.

The Fibonacci chain includes sub-chains of alternating $L$ and $S$ bonds that favor alternating angles and repeated $L$ bonds that favor  full polarization.
We consider the two possible sub-chains of the Fibonacci chain $LLSLL$ and $LLSLSLL$, in which the sequence $LL$ is found only at the ends. Assuming that the spins in the middle of consecutive $L$ bonds are parallel to the $\hat{z}$-axis, we can deduce the expected angles for the other spins in the sub-chain by minimizing the energy. For the $LLSLL$ sub-chain, the spin to the left of the $S$ bond has an angle $\phi_{a} \simeq (D_L-D_S)/(H-J_L-J_S)$, and the one to the right of the $S$ bond has an angle $-\phi_{a}$.
For the $LLSLSLL$ sub-chain, two angles are involved:
\begin{equation}
\phi_{b1} \simeq \frac{(D_L-D_S)(H-2J_L)}{H^2-3H J_L+2J_L^2-2 HJ_S + 3J_L J_S},
\end{equation} and
\begin{equation}
\phi_{b2}\simeq \frac{H-2J_L}{H-J_L}\phi_{b1},
\end{equation}
that correspond to the second and third spins from left to right in the sub-chain, respectively. The forth and fifth spins are parameterized with angles $-\phi_{b1}$ and $-\phi_{b2}$, respectively.
By using these results we can calculate the magnetization
\begin{equation}
   M = \left|\sum_{i=1}^{N}\vec{S}_i\right|,
\end{equation}
by counting the number of times these two sub-chains and $LL$ appear in the Fibonacci chain. The number of spins in the middle of $LL$ of bonds tends to $N\tau^3/(1+2\tau)$ in the large $N$ limit~\cite{baake2013aperiodic}. 
The number of  spins with the $\pm \phi_a$ angles is twice the number of $LLSLL$ sub-chains: $2 N/\tau^5$. The number of  spins with angles $\pm \phi_{b1}$ and   $\pm \phi_{b2}$  is equal to twice the number of $LLSLSLL$ sub-chains: $2 N/\tau^4$. The resulting magnetization is
\begin{equation}
    \frac{M_{\rm QFP}}{M_{\rm sat}}\simeq \frac{1}{1+2\tau}+\frac{2\cos(\phi_a)}{3+5\tau}+\frac{2\cos(\phi_{b1})+2\cos(\phi_{b2})}{2+3\tau}.
\end{equation}
where the saturation value is $M_{\rm sat}=N$.
This expression provides an accurate approximation of the magnetization in the QFP phase. Further improvements can be systematically achieved by considering larger sub-chains that span the Fibonacci chain for the angle minimization.

\subsection{Chiral solitons on a quasi-fully polarized background}
\begin{figure}[t]
\includegraphics[width=\columnwidth]{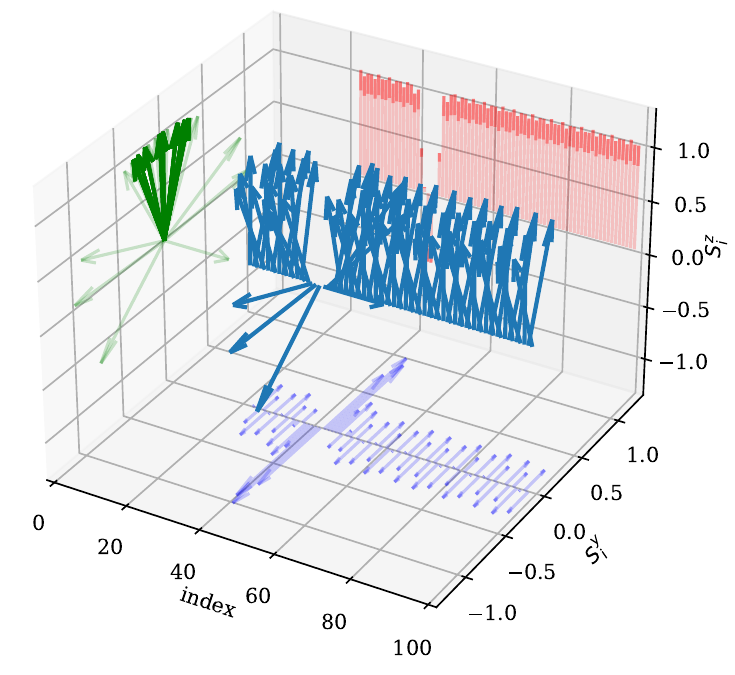}
\caption{Single-soliton configuration at $H=0.75$. Other parameters as in Fig. \ref{fig:3dstates_helical}. 
}
\label{fig:3dstates_soliton}    
\end{figure}

Before delving into the analysis of chiral soliton lattices, it is instructive to initially characterize isolated solitons on top of the QFP state, as depicted in Fig.~\ref{fig:3dstates_soliton}.
The QFP phase is the global ground state for $H \geq H_c$ and it remains as a metastable solution for $H \lesssim H_c$. The single-soliton solution is stable at $H=H_c$, but it remains as a metastable solution over a 
 finite interval of magnetic fields around $H=H_c$.
\begin{figure}[t]
\includegraphics[width=\columnwidth]{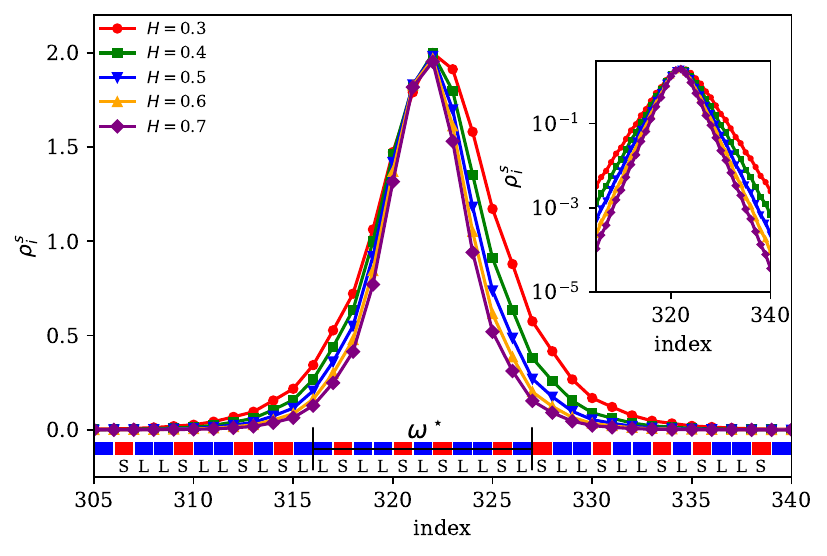}
\caption{Isolated soliton on the Fibonacci chain.  The change in the spin configuration $\rho^s_i$ produced by the soliton is shown for different magnetic field intensities.  The bond configuration is indicated using blue ($L$) and red ($S$) segments.  As it can be seen in the inset, $\rho^s_i$ decays exponentially away from the soliton center with a characteristic length $l_s$ that increases with decreasing magnetic field. Other parameters as in Fig. \ref{fig:3dstates_helical}.}
\label{fig:1soliton}    
\end{figure}
Figure \ref{fig:1soliton} shows the magnitude of the  single-soliton distortion relative to the QFP state:
\begin{equation}
\rho^{s}_i(H) = |\vec{S}^s_i(H)-\vec{S}^{\rm QFP}_i(H)|,
\end{equation}
at several magnetic fields for which both solutions are metastable. 
$\vec{S}^{\rm QFP}_i(H)$ is the quasi-fully polarized spin configuration and $\vec{S}^s_i(H)$ is the single-soliton  solution (see Fig.~\ref{fig:3dstates_soliton}). 

The inset of Fig. \ref{fig:1soliton} depicts the exponential decay of $\rho^s_i$ away from the soliton center, denoted as $x_s=\sum_i i\rho^s_i/ \sum_i\rho^s_i $. This behavior mirrors that observed in the homogeneous case, where the soliton size also diminishes with increasing magnetic field.

The lower portion of Figure \ref{fig:1soliton} delineates the bond configuration, highlighting the presence of the eleven-bond sub-chain $\omega^\star=LSLLSLSLLSL$ positioned at the core of the soliton. Across various parameter sets represented in the figure, this distinctive arrangement serves as a hallmark feature characterizing all locally stable single-soliton states for $H\lesssim H_c$. As solitons exhibit localized behavior with exponentially decaying tails, their local energy minima in the Fibonacci chain are determined by a finite sub-chain of bonds. While the precise composition of this sub-chain varies with model parameters, its identification enables pinpointing local stability positions for each instance within the Fibonacci chain.

For a concrete example, our analysis will focus on the parameter set outlined in Fig.~\ref{fig:1soliton}~\footnote[3]{This selection of parameters, where the DMI couplings are comparable to the exchange couplings, enables us to investigate the impact of DMI on ground state configurations while maintaining a manageable system size. The qualitative findings remain consistent for alternative parameter sets where $J_S, J_L<0$ and $D_S$, $D_L$ share the same sign.}. This choice yields the sub-chain $\omega^\star$. However, the methodology outlined below is adaptable to various other scenarios.

Due to the structure of the Fibonacci chain, the locations of the shorter sub-chain $\omega_0=SLS$ in the middle of $\omega^\star$ determine the locations of $\omega^\star$.  In fact, $\omega^\star$ is the only 11-bond extension of $\omega_0$, allowed in the Fibonacci chain, that can be obtained by attaching an equal number of bonds (though not necessarily identical ones) on each side of $\omega_0$. 

In a Fibonacci chain, a specific sub-chain, such as  $\omega_0$, reappears at intervals given by Fibonacci numbers~\cite{huang2015sequence}. Specifically, the number of bonds separating the centers of two consecutive instances of a sub-chain $\omega_s$ is one of two consecutive Fibonacci numbers denoted $I_{L}$, and $I_{S}$, where $I_{L} > I_{S}$.  ($I_L=8$ and $I_S=5$ for the sub-chains $\omega_0$ and $\omega^\star$).  These intervals follow the same pattern as the sequence of bonds in the Fibonacci chain, i.e. making the replacement $L\to I_L$, $S\to I_S$:   
$$LSLLSLSL\ldots\to I_{L} I_{S}I_{L}I_{L}I_{S}I_{L}I_{S}I_{L}\ldots. $$  
It is important to keep in mind  that the length of a sub-chain $|\omega_s|$ -the count of bonds it comprises-  may me larger than the distance to its nearest identical sub-chain, because two consecutive instances of a sub-chain may have a finite overlap (share several bonds). For example the sub-chain $\omega^\star$ has a length $|\omega^\star|=11 >I_S=5$.  However, the minimum distance $I_S$ between instances of a sub-chain $\omega_s$ is of the order of the size of the sub-chain itself $I_S\sim|\omega_s|$.

Owing to the quasiperiodic structure of the Fibonacci chain, each point within the chain is distinct. Consequently, every instance of $\omega_0$ is embedded in a unique bond ``environment''. As a result, the energy associated with a soliton located at these different instances of $\omega_0$ will necessarily vary, reflecting the distinct local environments within the chain. 

We define the energy of a single soliton at a specific position $x_s$ in the chain as the energy difference 
\begin{equation}
   \varepsilon_s(x_s)=E_{\rm 1s}(x_s)-E_{\rm QFP}.  
\end{equation}
between the state with a single soliton centered at $x_s$ and the QFP state.

Analyzing the impact of the surrounding environment on soliton energy entails investigating all extensions of the central sub-chain obtained by appending an equal number of bonds to each side. Variations in the sub-chains, starting from $\omega_0=SLS$, only occur for extensions of 13 bonds, precisely at the sixth bond away from the center of the sub-chain. Table \ref{tab:environments} provides a breakdown of the three potential 13-bond extensions.
\begin{table}[t]
    \centering
\begin{tabular}{c|c|c}
   Name  & Bond configuration &Size   \\
   \hline
   \hline
    $\omega_0$& $SLS$&3\\
   \hline
     $\omega_{1b}$  &$W_\alpha^S\oplus\omega_0\oplus W_\beta^S$ &\\
     $\omega_{1a}$  &$W_\alpha^S\oplus\omega_0\oplus W_\alpha^S$ &13\\ 
     $\overline{\omega}_{1b}$&$\overline{W}\vphantom{W}_\beta^S\oplus\omega_0\oplus W_\alpha^S$&\\
   \hline
     $\omega_{2c}$ &$ W_\alpha^L\oplus\omega_{1b}\oplus W_\beta^L$&\\
      $\omega_{2b}$&$W_\beta^L \oplus\omega_{1b}\oplus W_\beta^L$&\\
    $\omega_{2a}$&$W_\alpha^L\oplus\omega_{1a}\oplus W_\alpha^L$& 29\\
      $\overline{\omega}_{2b}$&$ \overline{W}\vphantom{W}_\beta^L\oplus\overline{\omega}_{1b}\oplus \overline{W}\vphantom{W}_\beta^L$&\\
      $\overline{\omega}_{2c}$&$\overline{W}\vphantom{W}_\beta^L\oplus\overline{\omega}_{1b}\oplus W_\alpha^L$&\\
\end{tabular}
    \caption{Allowed extensions of $\omega_0=SLS$ up to the second generation. Here $W_\alpha^S=LLSLL$, $W_\beta^S=LLSLS$,  $W_\alpha^L=SLSLLSLS$, and $W_\beta^L=LLSLLSLS$.}
    \label{tab:environments}
\end{table}

There is a inversion symmetric sub-chain $\overline{\omega}_{1a}=\omega_{1a}$, where the overline indicates an inversion in the bond order of the sub-chain.  This extension is obtained attaching the 5-bond chain  $W_\alpha^S=LLSLL$ on each side of $\omega_0=SLS$: $\omega_{1a}=W_\alpha^S\oplus\omega_0 \oplus W_\alpha^S$.
The additional sub-chains, $\omega_{1b}=W_\alpha^S\oplus \omega_0\oplus W_\beta^S$ and its inversion partner $\overline{\omega}_{1b}=\overline{W_\alpha^S\oplus \omega_0\oplus W_{\beta}^S}=\overline{W
}\vphantom{W}_\beta^S\oplus \overline{\omega}_0\oplus \overline{W}\vphantom{W}_\alpha^S=\overline{W
}\vphantom{W}_\beta^S\oplus \omega_0\oplus W_\alpha^S$ can be constructed by using   $W_\beta^S=LLSLS$ and $\overline{W}\vphantom{W}_\beta^S$.

These diverse local environments result in varying soliton energies. However, it is crucial to account for an important symmetry. The Hamiltonian remains invariant under a chain inversion coupled with a mirror reflection on the $xz$ plane. Consequently, soliton energies at locations that are inversion symmetric, such as $\omega_{1b}$ and $\overline{\omega}_{1b}$, are not distinguished by an environment of that size.
The primary energy splitting, labeled as $\varepsilon_1$, occurs between the inversion symmetric $\omega_{1a}$ and the pair $(\omega_{1b}, \overline{\omega}_{1b})$. A similar pattern for the energy splitting is expected for larger environments.

\begin{figure}
    \centering
    \includegraphics[width=\columnwidth]{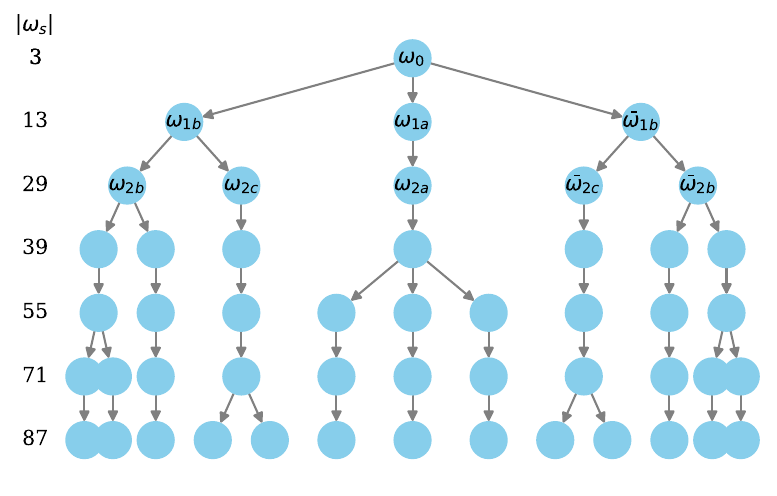}
    \caption{Soliton environment genealogy of the sub-chain $\omega_0=SLS$.}
    \label{fig:tree}
\end{figure}

The next difference in the environment is obtained for 29-bond sub-chains that are obtained after attaching 8-bond chains $W_\alpha^L=SLSLLSLS$, $W_\beta^L=LLSLLSLS$, or $\overline{W}\vphantom{W}_\beta^L$ to each side of the sub-chains of the previous generation: $\left\{\omega_{1a},\, \omega_{1b},\, \overline{\omega}_{1b}\right\}$. Only five of the 29-bond extensions constructed in this way are allowed. This introduces an energy splitting $\varepsilon_2$ for the pair $(\omega_{1b}, \overline{\omega}_{1b})$. Due to the localized nature of the soliton states, the splitting generated at each new generation of derived sub-chains is expected to decrease exponentially in the length of the sub-chain, 
as it is caused by bonds further away from the soliton center.

This behavior recurs in a quasiperiodic manner. To generate the next set of sub-chains, either the 5-bond set $W^S=\left\{ W^S_\alpha, W^S_\beta, \overline{W}{\vphantom{W}^S_\beta}\right\}$ or the 8-bond set $W^L=\left\{ W^L_\alpha, W^L_\beta, \overline{W}{\vphantom{W}^L_\beta}\right\}$ must be attached to the sub-chains from the previous generation. Only resulting sub-chains present in the Fibonacci chain are considered. Attachments leading to two consecutive $S$ bonds or four consecutive $L$ bonds are disregarded.
Each new generation increases the number of sub-chains by two (a sub-chain and its inversion partner), breaking the degeneracy either between a pair of inversion partners or the symmetric extension, as depicted in Fig.~\ref{fig:tree}.

For $\omega_0=SLS$, the size of the chains to be attached to create a new generation of environments follows a Fibonacci sequence.  After the first step in which the 5-bond chains $W^S$ are used, the subsequent extensions  can be obtained reading the Fibonacci sequence $W^LW^SW^LW^LW^SW^LW^S\ldots$, from left to right. The first few generations of sub-chains created from the root sub-chain $\omega_0=SLS$ are presented in Fig. \ref{fig:tree}. This family tree presents a mirror symmetry about a vertical axis that goes through the node labeled $\omega_0$. All sub-chains associated with nodes on this axis are inversion symmetric. Nodes connected by the mirror symmetry correspond to inversion partners.

This sequence establishes a hierarchical structure in the single-soliton energies, as the energy difference between two locations with varying bond configurations solely for sub-chains of length $\geq |\omega_s|$ is anticipated to be proportional to $e^{-\alpha|\omega_s|/2l_s}$, where $\alpha\approx 1$. In other words, it is expected to exponentially decrease with the distance from the soliton center to the first differing bond.

The bond structure of the Fibonacci chain shapes an energy landscape for the solitons with local minima that present a singular distribution of energy and follow a aperiodic pattern. This spectrum bears resemblance with the one obtained for tight-binding models on the Fibonacci chain \footnote[2]{see Ref. \onlinecite{Jagannathan21} and references therein.}.

\begin{figure*}[t]
\includegraphics[width=2\columnwidth]{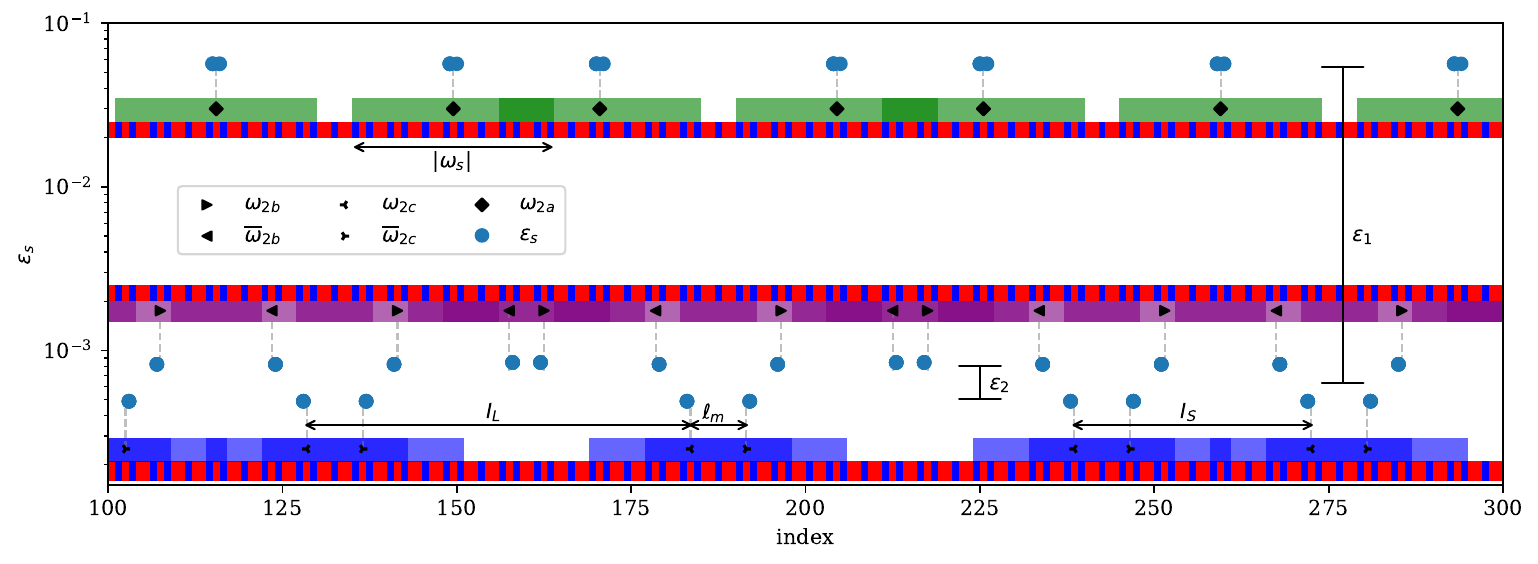}
\caption{Energy of a single soliton, $\varepsilon_s$, at stable positions in the Fibonacci chain under a magnetic field $H=0.63925$, exceeding the critical field $H_c \approx 0.63915$. Positions corresponding to the 29-bond sub-chains from Table \ref{tab:environments} are marked with symbols and rectangles, vertically offset for clarity. Darker shades in overlapping rectangles signify shared bonds between sub-chains. The Fibonacci chain's bond arrangement is also depicted. Other are parameters as in Fig. \ref{fig:3dstates_helical}. }
\label{fig:soliton}    
\end{figure*}
The position dependent soliton energy  and its relation to the bond environment are presented in Fig.~\ref{fig:soliton} for a magnetic field marginally larger than  $H_c$. The centers of the 5 different 29-bond sub-chains stemming from $\omega_0=SLS$ are indicated with different symbols and their span is indicated with rectangles. Darker shades in overlapping rectangles signify shared bonds between sub-chains. 
 
The figures also shows the minimum distance $\ell_m$ between a sub-chain and its inversion symmetric counterpart.
The two possible distances between consecutive identical sub-chains, $I_S$ and $I_L$, are also highlighted along with the two relevant  energy level splittings $\varepsilon_1$ and $\varepsilon_2$.

The sub-chains $\omega_{2c}$ and $\overline{\omega}_{2c}$ represent the two lowest lying energies and are anticipated to split only in the 87-bond sub-chain generation, as illustrated in Fig. \ref{fig:tree}. In simpler terms, these locations exhibit an identical sequence of bonds up to the 43rd bond away from the center.

\subsection{Chiral soliton lattices}

For magnetic fields exceeding the critical field $H_c$, the system's ground state transitions to the non-collinear forced QFP phase as described previously. As depicted in Fig. \ref{fig:soliton}, metastable  single-soliton states  with energy $\varepsilon_s$ emerge within a finite range of fields above $H_c$. Near $H_c$, the magnetic field works as a  chemical potential for the solitons, implying that $\varepsilon_s$ is linear in $H-H_c$. For $H$ slightly lower than $H_c$ the single-soliton energy becomes negative  for solutions centered around specific sub-chains (``pinning centers'') of the  Fibonacci chain.

\begin{figure}
    \centering
    \includegraphics[width=\columnwidth]{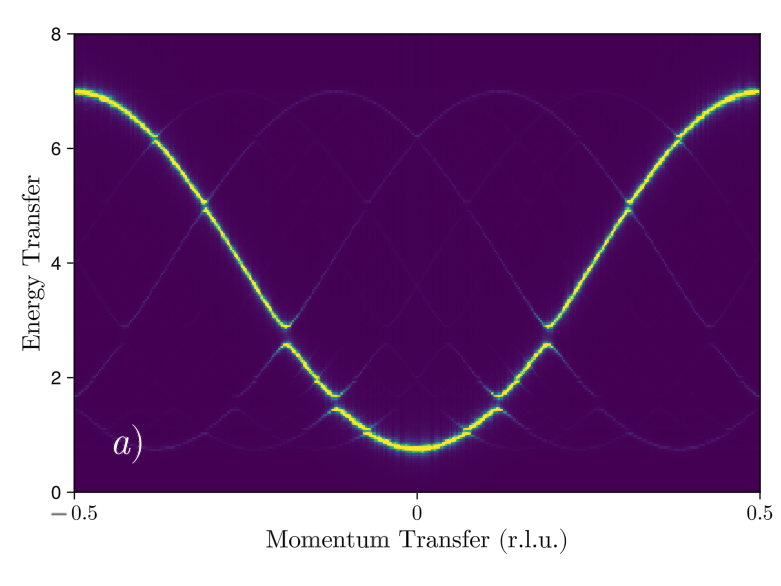}
    \includegraphics[width=\columnwidth]{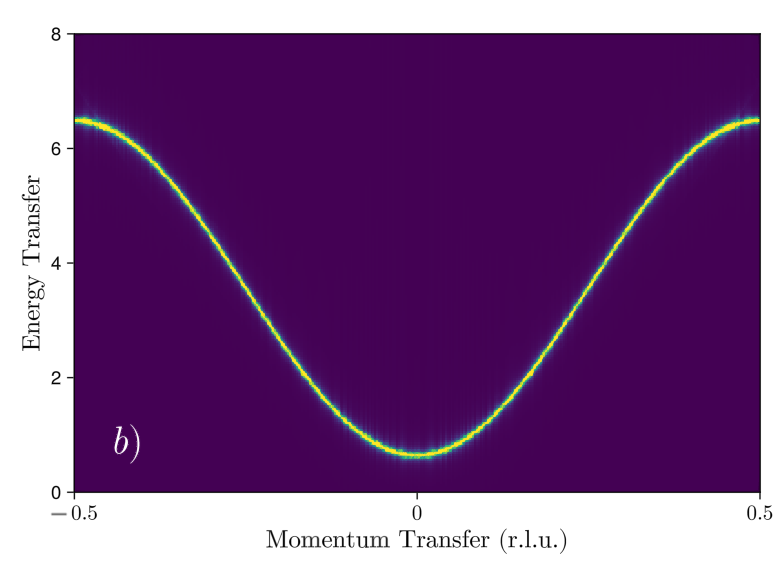}
    \caption{Dynamic spin structure factor $S(q,\omega)$ for a 377 site chain with periodic boundary conditions at $H\simeq H_c$, where $H_c$ is the critical field of the Fibonacci chain, and $k_B T =0.0005$.  The numerical calculations were performed using the Sunny package~ \cite{dahlbom2022geometric,dahlbom2022langevin,dahlbom2024quantum}. a) Fibonacci chain with couplings $J_S=-1.0 $, $J_L=-1.75$, $D_S=-2.25$, and $D_L=-0.75$. 
    b) Homogeneous chain with first neighbor couplings given by the Fibonacci chain average values $\langle J\rangle=J_S/\tau^2+ J_L/\tau \approx -1.464 $ and $\langle D\rangle=D_S/\tau^2+ D_L/\tau  \approx -1.323$.} 
    \label{fig:Sq}
\end{figure}

An intriguing aspect of this \emph{continuous} field-induced phase transition is the retention of a gapped spin wave spectrum at the critical field $H=H_c$. In essence, the softening of chiral solitons does not coincide with a softening of the spin waves. This situation, 
which holds true regardless of the periodic or quasi-periodic nature of the model,
is reminiscent of field-induced multipolar orderings where an $n$-magnon ($n \geq 2$) bound state becomes gapless, while single magnon modes remain gapped. Indeed, as shown in Fig.~\ref{fig:Sq} the spin wave gap at $H=H_c$ is $\Delta_s \approx 0.74$ for the Fibonacci chain and $\Delta_s \approx 0.64$ for the homogeneous chain. Due to the gapped nature of the spin wave spectrum, the chiral soliton exhibits a characteristic size $\sim l_s$, and the interaction between solitons separated by a distance $l$ diminishes exponentially with $l/l_s$.

Since negative-energy locations are typically separated by distances much larger than $l_s$ just below $H_c$, one might intuitively anticipate that all locations with negative single-soliton energy would be occupied by a soliton. However, two subtleties challenge this simplistic reasoning. Firstly, the energy difference between ``pinning centers'' diminishes exponentially with the minimum sub-chain size $|\omega_s|$, necessary to differentiate between the two centers. Consequently, if the distance between pinning centers is comparable to $|\omega_s|$, only one of the centers may be occupied over a small field range.

The second subtlety is that for a non-symmetric sub-chain $\omega_s$ ($\overline{\omega}_s\neq\omega_s$) the reverse and quasi-degenerate partner $\overline{\omega}_s$ is at a distance $\ell_m <|\omega_s|/2$  ( $|\omega_s|/\tau\leq \ell_m \leq|\omega_s|\tau$ for symmetric sub-chains). This means that the repulsive interaction energy can be stronger than the potential energy gain.

\subsubsection{Fields near the critical field \texorpdfstring{($H\lesssim H_c$)}{(H\~ Hc)}}
To make the analysis more quantitative, we choose a magnetic field $H$  such that the energy of the two low lying pinning centers around $\omega_s$ and $\overline{\omega}_s$ is $\varepsilon_s=-\varepsilon/2$ (we neglect smaller energy differences for the moment), while the excitation energy ($\varepsilon_2$ in Fig.~\ref{fig:soliton}) to the next level ($\omega_{2b}$ and $\overline{\omega}_{2b}$  in Fig.~\ref{fig:soliton})  is $\varepsilon$, implying that the corresponding sub-chain locations are unstable pinning centers.
 The energy difference between occupying both  consecutive minima at $\omega_s$ and $\overline{\omega}_s$ and only one is then 
\begin{equation}
\epsilon_I - \frac{\epsilon}{2} = I_0 e^{-\gamma\ell_m/l_s} - \frac{\epsilon_0}{2} e^{-\alpha|\omega_s|/2 l_s},
\label{Eq:ce}
\end{equation}
where $\epsilon_I= I_0 e^{-\gamma\ell_m/l_s}$ is the interaction energy between two solitons. We find numerically for the coupling parameters $J_S=-1.0 $, $J_L=-1.75$, $D_S=-2.25$, and $D_L=-0.75$, that $I_0$ is an order of magnitude larger than $\epsilon_0$, $l_s\approx 1.52$, $\gamma\approx 0.92$, and $\alpha\approx 0.85$. 
For $H\to H_c$, $|\omega_s|\to \infty$,  given that  $\ell_m< |\omega_s|/2$ the quantity in  Eq. (\ref{Eq:ce}) is expected to be positive and only one of two nearest neighbouring locations $\omega_s$ or $\overline{\omega}_s$ to be occupied.

To determine the magnetization of each generation of soliton lattices near the saturation field, we introduce $\Delta M = M_{\rm QFP} - M_{\rm sol}$, where $M_{\rm sol}$ is the magnetization of the single-soliton solution.  The pair of conjugate sub-chains at distance $\ell_m$ appear quasiperiodically following the Fibonacci sequence for $I_S$ and $I_L$.  The above-described solutions have one soliton in each pair of conjugate minima $\omega_s$ and $\overline{\omega}_s$. Therefore, there is one soliton per sub-chain $\omega_s$, implying that the total number of solitons is $N_L + N_S$, where $N_L$ ($N_S$) is the number of $I_L$ ($I_S$) segments:  

\begin{align} \label{eq:mag}
\frac{M}{M_{\rm sat}} &= \frac{M_{\rm QFP}}{M_{\rm sat}} - \frac{(N_L+N_S)\Delta M}{M_{\rm sat}} \nonumber \\
&= \frac{M_{\rm QFP}}{M_{\rm sat}}  - \frac{\left( 1+ \frac{N_L}{N_S} \right) }{\left( 1 + \frac{N_L}{N_S} \frac{I_L}{I_S}   \right)} \frac{\Delta M}{I_S}
\nonumber \\
&\simeq \frac{M_{\rm QFP}}{M_{\rm sat}}  - \frac{\left( 1+ \tau \right) }{\left( 1 + \tau \frac{I_L}{I_S}   \right)} \frac{\Delta M}{I_S}
\nonumber \\
&\simeq \frac{M_{\rm QFP}}{M_{\rm sat}} - \frac{\left( 1+ \tau \right) }{\left( 2 + \tau   \right)} \frac{\Delta M}{I_S} ,
\end{align}
which becomes exact for $|\omega_s| \to \infty$. Note that $M/M_{\rm QFP}$ becomes asymptotically close to $1$ when $|\omega_s| \to \infty$. This behavior occurs because $I_S$, which is lower bounded by $|\omega_s|/\tau^2$, tends to infinity under these conditions.

To obtain the ground state magnetic configuration right below $H=H_c$, the magnetic energy given by Eq.~(\ref{eq:totenergy}) needs to minimized for all the angles $\phi_i$ that characterize the orientation of a single spin in the plane perpendicular to the helimagnet axis.
Near the saturation field $H_c$, however, the following procedure can be applied: 
\begin{enumerate}
\item Identify the locations on the Fibonacci chain where the soliton energy is negative. A sub-chain $\omega_s$, and eventually its inversion partner $\overline{\omega}_s$, common to all of these location is determined. As discussed earlier regarding the structure of the Fibonacci chain, the size of these sub-chains, represented by $|\omega_s|=|\overline{\omega}_s|$, expands as the difference $H_c-H$ decreases. 
\item Determine the relevant distances between instances of $\omega_s$ and $\overline{\omega}_s$: $\ell_m$, $I_S$, $I_S-\ell_m$, $I_S+\ell_m$, $I_L-\ell_m$, and $I_L+\ell_m$. %
\item Compute the interactions between solitons separated by these distances. The interaction energy between solitons at positions $x_{1}$ and $x_2$, with corresponding energies $\varepsilon_{1s}(x_1)$ and $\varepsilon_{1s}(x_2)$, can be obtained from the energy of the two soliton state $E_{2s}$ as.
\begin{equation}
V(x_{1}-x_{2}) = E_{2s}(x_1,x_2)-2\varepsilon_{s}-E_{\rm QFP},
\end{equation}
where we have used that the single-soliton energies are quasidegenerate $\varepsilon_{1s}(x_1) \approx \varepsilon_{1s}(x_2)\approx \varepsilon_s<0$.
\item  Obtain the distance  $x_{min}=|x_1-x_2|$ defined as the minimum separation at which $V(x_1-x_2)+\varepsilon_s$ becomes negative. This distance determines which soliton pairs can coexist in the ground state.
\item Place solitons in negative-energy locations. Starting with one soliton and progressively adding other  at the closest possible locations that are at a distance larger than $x_{min}$. 
\item 

The final step involves addressing situations in which a change in the location of a soliton to an adjacent pinning center does not modify the interaction energy with its nearest neighbors. This entails investigating whether relocating a soliton to an adjacent negative-energy location alters the distances to its nearest neighboring solitons in the chain. This scenario occurs if the distance to the left soliton is denoted as $d_a$, the distance to the right soliton as $d_b$, and moving the central soliton results in swapping $d_a$ and $d_b$. To resolve this degeneracy, a thorough analysis of the larger environments surrounding the two available position spots is conducted, extending beyond the sub-chain $\omega_s$. Once a criterion for resolving this degeneracy is established, it can be uniformly applied to all similar instances recurring in a quasiperiodic manner.
\end{enumerate}

\begin{figure}
   \includegraphics*[width=\columnwidth]{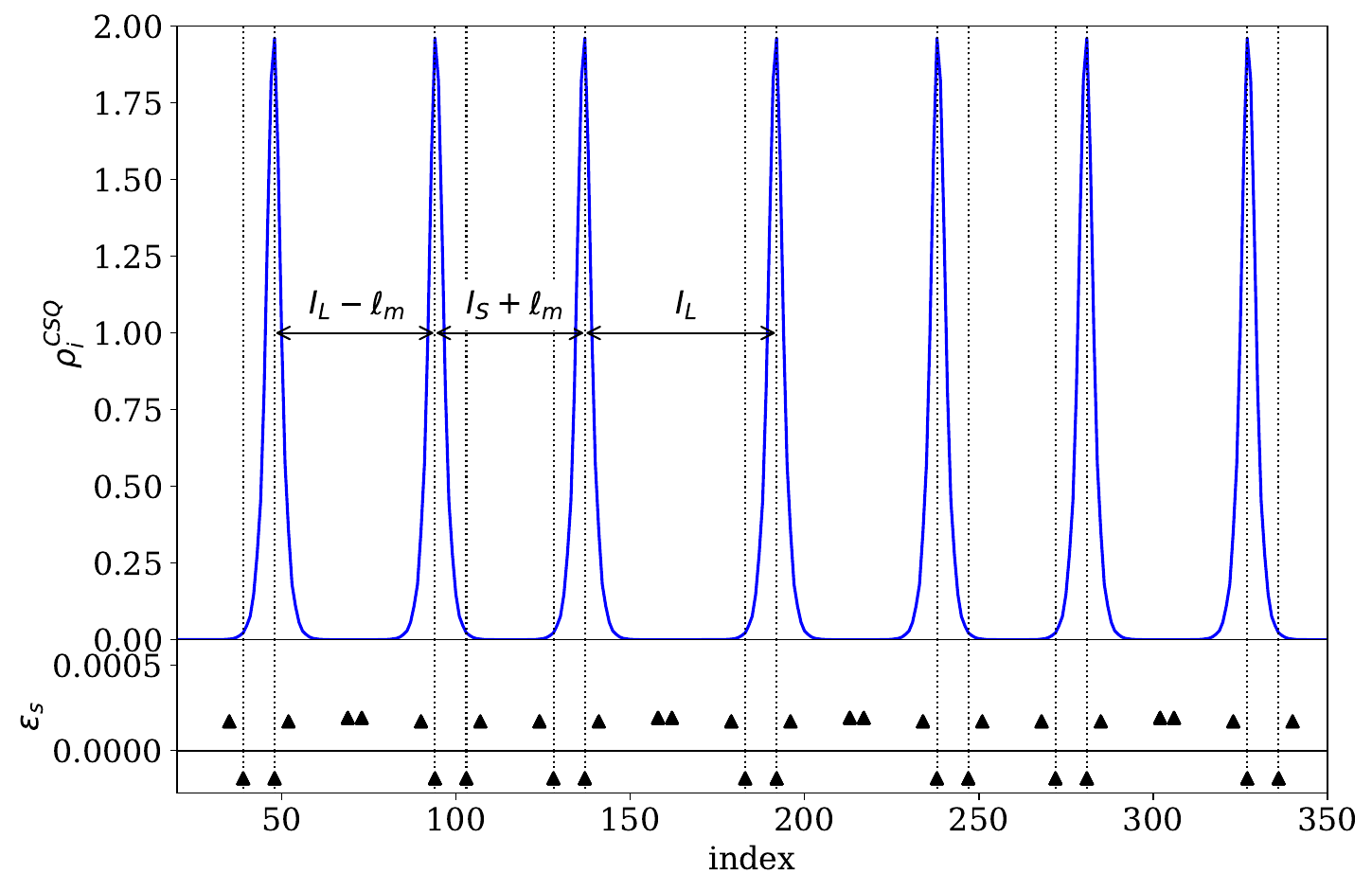}
    \caption{The top panel presents the ground state configuration for an external magnetic field $H=0.6375$. The lower panel presents the lowest single-soliton energies as a function of spin index. The vertical lines indicate negative-energy single soliton locations. Other parameters as in Fig. \ref{fig:3dstates_helical}. }
   \label{fig:2dplot12783}
\end{figure}

This magnetic field regime is exemplified in Fig. \ref{fig:2dplot12783} where the ground state configuration for a magnetic field $H=0.63915\simeq H_c$  is shown. More specifically, the quantity plotted is
\begin{equation}
\rho^{CSQ}_i = |\vec{S}^{CSQ}_i-\vec{S}^{QFP}_i|
\end{equation}
where $\vec{S}^{CSQ}_i$ is the ground state spin configuration corresponding to a chiral soliton quasicrystal (CSQ).
The single-soliton energies and their locations are also indicated in the figure. The position of the solitons on the lattice correspond to the negative-energy spots, as indicated in the lower panel of Fig. \ref{fig:2dplot12783}, where the spots that correspond to the sub-chains $\omega_{2c}$ and $\overline{\omega}_{2c}$ ($|\omega_{2c}|=29$) exhibit a negative energy $\varepsilon_s$. Key distances relevant to the soliton positions include $I_L=55$, $I_S=34$, $\ell_m=9$, $I_L-\ell_m=46$, $I_S-\ell_m=25$, $I_S+\ell_m=43$, and $I_L+\ell_m=64$. Through numerical analysis, the minimal distance for optimal soliton placement is determined to be $x_{min}=I_S+\ell_m$. Consequently, only one soliton occupancy is permissible for pairs of sub-chains ($\omega_{2c}$, $\overline{\omega}_{2c}$) separated by a distance $\ell_m$. The spacing between these pairs adheres to the Fibonacci quasiperiodic sequence observed in $I_L$ and $I_S$. When pairs are separated by $I_S$, the solitons are positioned at a distance $I_S+\ell_m$, thereby occupying the two external spots. In cases where pairs are separated by $I_L$, and if flanked by pairs at a distance $I_S$, the inner spots of the $I_L$ pair are filled. However, if one side features another $I_L$ separation, an interaction degeneracy arises concerning the occupancy of the pair of spots within the middle of the $I_LI_L$ sequence. This degeneracy is effectively resolved by examining the surrounding environment of these two spots until a point where differentiation within the sub-chains occurs, in this case at 87-bond sub-chains. The resulting energy splitting is exceptionally small (approximately on the order of $10^{-11}$).

\begin{figure}
    \centering
    \includegraphics*[width=\columnwidth]{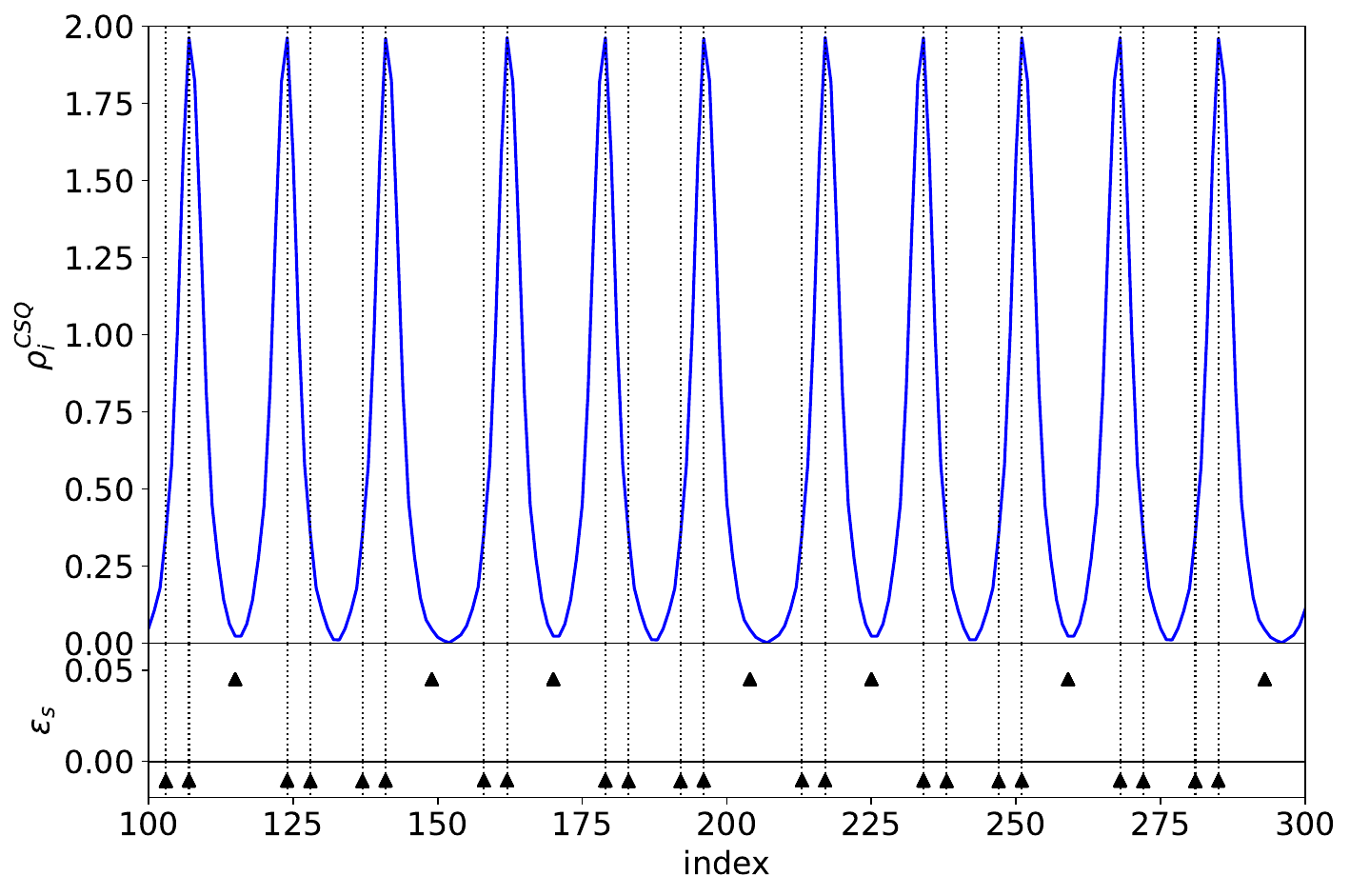}
    \caption{Same as Fig. \ref{fig:2dplot12783} for $H=0.635$.}
        \label{fig:2dplot1275}
\end{figure}

As shown in Fig. \ref{fig:2dplot1275}, a similar situation arises for a lower magnetic field. The negative soliton energy spots in this case are associated with the 13-bond sub-chains $\omega_{1b}$ and $\overline{\omega}_{1b}$. Here, there is also an interaction degeneracy, which is lifted by the energy splitting that occur for 29-bond sub-chains.

When this type of degeneracies are lifted uniformly for all instances along the Fibonacci chain, the quasicrystalline soliton structure can readily constructed. The starting point are the chains
\begin{align}
   \tilde{\Omega}_1 &= \Omega(I_L)\oplus \Omega(I_L)\oplus \Omega(I_S)\nonumber \\
   \tilde{\Omega}_2 &=  \Omega(I_L) \oplus \Omega(I_L)\oplus \Omega(I_S)\oplus \Omega(I_L)\oplus \Omega(I_S),\nonumber
\end{align}
where $\Omega(I_L)$ and $\Omega(I_S)$ indicate the sequences of $I_L$ and $I_S$ bonds, respectively, that separate occurrences of the chain $\omega_s$.
The soliton locations in these chains are uniquely determined and the Fibonacci concatenation rule $\tilde{\Omega}_m=\tilde{\Omega}_{m-1}\oplus\tilde{\Omega}_{m-2}$ can be used repeatedly to construct the quasicrystalline soliton lattice.

\subsubsection{`CDE' state}
As the magnetic field is reduced further, the single soliton energy $\varepsilon_s$ becomes negative for locations associated with the inversion symmetric sub-chain $\omega_{1a}$. The interactions between solitons prevent these locations to be occupied until the  energies $\varepsilon_s$ become negative enough to overcome the repulsive interactions. 

\begin{figure}
    \centering
    \includegraphics*[width=\columnwidth]{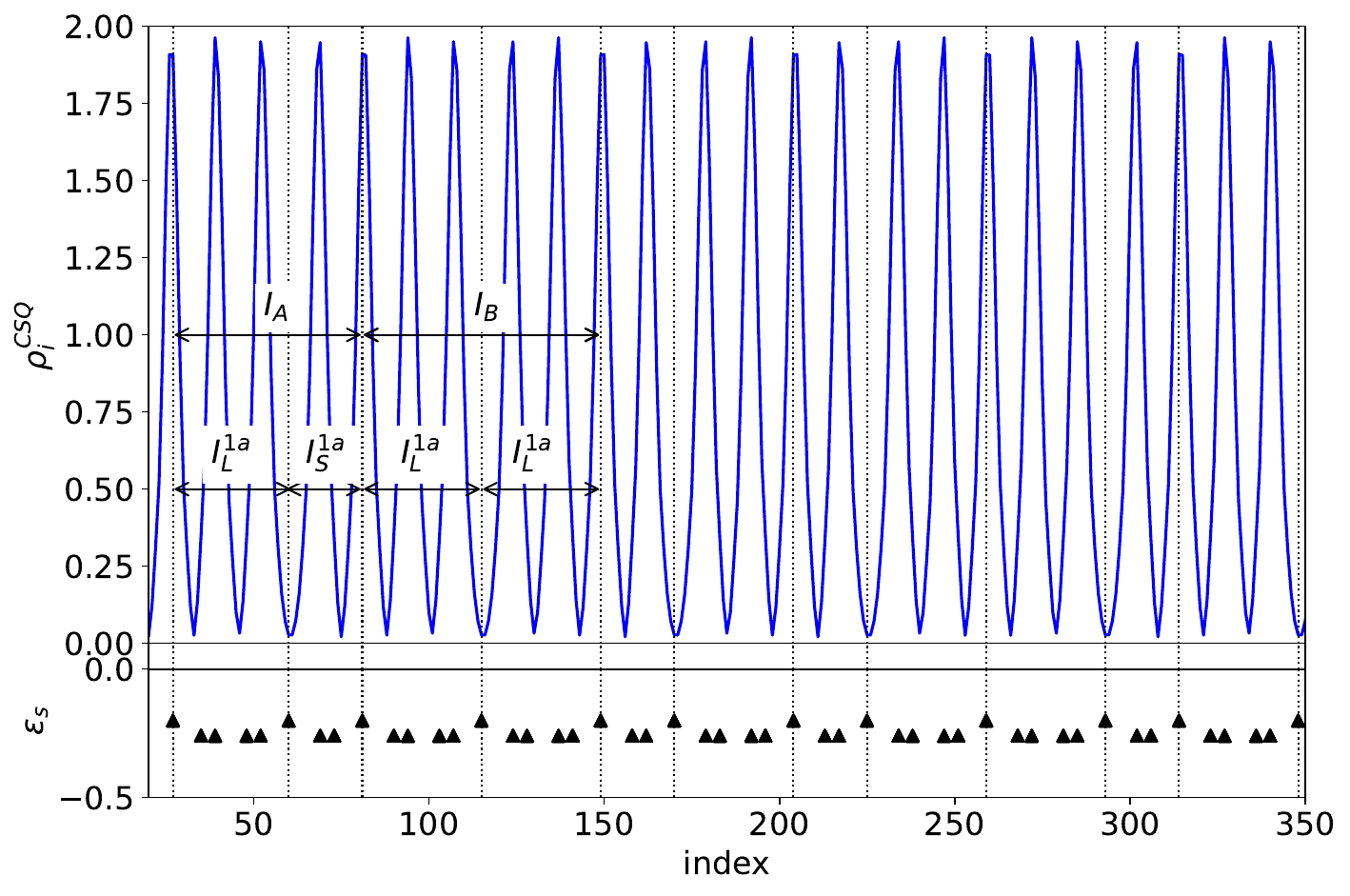}
    \caption{Same as Fig. \ref{fig:2dplot12783} for $H=0.6$.}
    \label{fig:2dplot12}
\end{figure}

As illustrated in Fig.~\ref{fig:2dplot12}, a commensurate chiral soliton lattice emerges wherein only half of the $\omega_{1a}$ locations are occupied by a soliton. This state is particularly simple because the quasi-degeneracy associated with $\omega_{1b}$ and $\overline{\omega}_{1b}$ pairs in the middle of an $I_LI_L$ interval is resolved due to the occupation of $\omega_{1a}$ locations. The interaction with the solitons on $\omega_{1a}$ locations renders the nearest neighboring $\omega_{1b}$ and $\overline{\omega}_{1b}$ spots unfavorable. This, in turn, also renders the nearest instances of $\omega_{1a}$ unfavorable.

In this state one out of two instances of $\omega_{1a}$ is occupied, making the distances between solitons in  $\omega_{1a}$ locations equal to  $I_A= I^{(1a)}_{L}+I^{(1a)}_S$ or $I_B=2I^{(1a)}_{L}$, where $I^{(1a)}_S=21$ and $I_L^{(1a)} =34$ are the distances between consecutive instances of $\omega_{1a}$. The sequence of $I_A$ and $I_B$ can be obtained following the expansion rule:
\begin{align}
    A&\to BAAA, \\
    B&\to BAAAA.
\end{align}
The inflation matrix relating the number of $A$ and $B$ at a given inflation step is
\begin{equation}
    \Bigg( \begin{matrix}
        N_A^{(n+1)} \\
        N_B^{(n+1)}
    \end{matrix} \Bigg) =
    \Bigg( \begin{matrix}
        3 & 4 \\
        1 & 1
    \end{matrix} \Bigg)
    \Bigg( \begin{matrix}
        N_A^{(n)} \\
        N_B^{(n)}
    \end{matrix} \Bigg)
\end{equation}
for which the largest eigenvalue $\tau^3$ is a Pisot-Vijayaraghavan number and the corresponding eigenvector determines the ratio $N_A/N_B=2\tau$ in the large-$N$ limit~\cite{Jagannathan21}.
The number of solitons in $\omega_{1a}$ locations is $N_A+N_B=\tau^3 N_B$ and the relative change in the magnetization is:
\begin{align}
    \delta_M&=-\Delta M\frac{N_A+N_B}{M_{\rm sat}}\nonumber \\
    &=-\Delta M\frac{N_A+N_B}{N_A I_A+ N_B I_B}\nonumber \\
    &=-\Delta M\frac{(2\tau+1)}{2\tau I_A+  I_B}\nonumber\\
    &=-\Delta M\frac{(2\tau+1)}{2\tau (I^{(1a)}_{L}+I^{(1a)}_S)+ 2 I^{(1a)}_{L}}\nonumber\\
    &=-\frac{\Delta M}{I^{(1a)}_S}\frac{(2\tau+1)}{(2\tau+2)\frac{I^{(1a)}_{L}}{I^{(1a)}_S}+ 2 \tau}\nonumber\\
    &\approx -\frac{\Delta M}{I^{(1a)}_S}\frac{(2\tau+1)}{(2\tau+2)\tau+ 2 \tau} 
    \nonumber \\
    &= -\frac{\Delta M}{I^{(1a)}_S}\frac{2\tau +1}{6\tau +2} =-\frac{1}{2}\frac{\Delta M}{I^{(1a)}_S}\frac{\tau^2}{2+\tau}.
    \end{align}
The magnetization of this state can be obtained using Eq. (\ref{eq:mag}) and the result for $\delta_M$:
\begin{align}
\frac{M}{M_{\rm sat}} &\approx \frac{M_{\rm QFP}}{M_{\rm sat}}  - \frac{\left( 1+ \tau \right) }{\left( 2 + \tau   \right)} \Delta M\left(\frac{1}{I_S^{(1b)}}+\frac{1}{2I_S^{(1a)}}\right),
\end{align}
where we have assumed that the change in the magnetization $\Delta M$ produced by a single soliton is the same for $\omega_{1a}$ and $\omega_{1b}$ locations.
 In this case we have $I_S^{(1b)}=13$ and $I_S^{(1a)}=21$.

\begin{figure}
\includegraphics[width=\columnwidth]{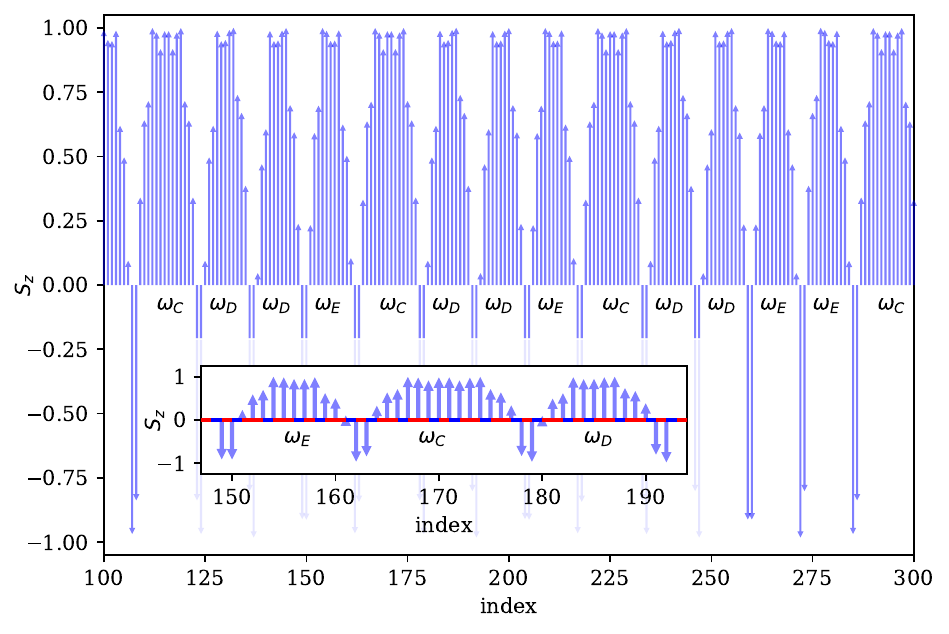}
\caption{Section of the ground state configuration for $H=0.6$. The projection of the spins along the $\hat{z}$-axis y represented using arrows.  $\omega_C$, $\omega_D$, and $\omega_E$ indicate sequences of bonds.  }\label{fig:CDEplot}
\end{figure}

The ground state spin configuration for $H=0.6$ is presented in Fig. \ref{fig:CDEplot}. The solitons separate three different types of domains in which the spins are predominantly aligned along the z-axis.  They correspond to three distinct sequences of bonds: $\omega_C=LSLLSLLSLSLLSLLS$, $\omega_D=LSLLSLSLLSLLS$, and the inverse $\omega_E=\overline{\omega}_D$.

The resulting sequence of $\omega_C$, $\omega_D$ , and $\omega_E$ that covers the Fibonacci chain can be obtained using the expansion rule
\begin{align}
C &\to DDEEC\nonumber \\
D &\to DDEC\\
E &\to DEEC \nonumber
\end{align}
starting from $C$. The largest eigenvalue of the inflation matrix is, as expected, the same as in the AB sequence above. The current sequence distinguishes the two possible orders of the short and long intervals ($I_LI_S$ and $I_SI_L$) that were not distinguished in the analysis of the soliton positions. The corresponding eigenvector indicates that the number of these segments satisfy $N_D=N_E$ and $N_C/N_D\to\tau$ in the thermodynamic limit.

\subsubsection{Lower fields}
When the magnetic field drops below approximately $\simeq 0.385$, there is a transition in the ground state configuration from the CDE state to another commensurate state where all $\omega_{1a}$ locations are occupied by solitons. Continuing the analysis that led to Eq.~\eqref{eq:mag} yields a magnetization
\begin{align}
\frac{M}{M_{\rm sat}} &\approx \frac{M_{\rm QFP}}{M_{\rm sat}}  - \frac{\left( 1+ \tau \right) }{\left( 2 + \tau   \right)} \Delta M\left(\frac{1}{I_S^{(1b)}}+\frac{1}{I_S^{(1a)}}\right),
\end{align}
for this state.

As the magnetic field diminishes within this phase, a critical field arises below which additional solitons that are not aligned with pinning centers emerge in the system. These extra solitons induce a discommensuration in the CSQ, mirroring the behavior observed in Frenkel-Kontorova models within a regular lattice \cite{braun2004frenkel}.

\section{Phase diagram} \label{sec:phasediagram}
\begin{figure}
\includegraphics[width=\columnwidth]{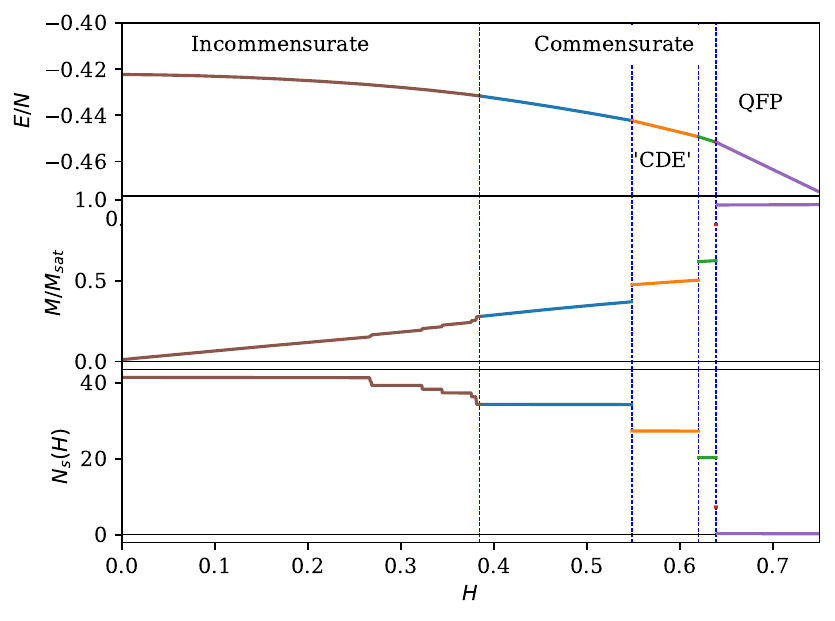}
\caption{Energy, magnetization and winding number as a function of the magnetic field for a Fibonacci chain of N=378 spins. Other parameters as in Fig. \ref{fig:3dstates_helical}.
}\label{fig:energy}
\end{figure}

The main findings of our study are depicted in Fig.~\ref{fig:energy}, which illustrates the ground state energy, magnetization, and winding number of a helimagnetic Fibonacci quasicrystal as a function of the external magnetic field.

Three main regimes have been identified: a quasi-fully polarized (QFP) state for high fields ($H>H_c$), a series of commensurate chiral soliton quasicrystalline lattices at intermediate fields ($H_{\rm CI}< H < H_c$), and incommensurate soliton lattices at low fields.

As we already mentioned, the high field QFP state is a non-colinear quasiperiodic spin configuration that can hold metastable chiral solitons. A so-called devil's staircase of commensurate quasicristalline lattices emerges right below the critical field $H_c$. The concept of commensurability in quasicrystals has been analyzed in the context of superconducting networks\cite{chaikin1988studies,grest1988commensurate,griffiths1992exact,itzler1994absence}. The existence of an inflation rule for the generation of the underlying quasicrystalline structure is not a necessary condition to produce commensurability effects \cite{griffiths1992exact,itzler1994absence}. However, when such rule is present, as in the Fibonacci quasicrystal under consideration, it offers an intuitive framework to analyze commensurability effects~\cite{chaikin1988studies}. For a commensurate lattice, we can find large enough Fibonacci subchains $\Omega_m$ and $\Omega_{m+1}$ that allow us to construct the soliton lattice using the concatenation rule of Eq.~\ref{eq:concat}.

In the Fibonacci helimagnet under consideration, the quasiperiodic nature of the couplings gives rise to an effective quasiperiodic potential energy for the solitons. The arrangement of soliton lattices at a specific magnetic field hinges on the interplay between soliton-soliton interactions and the underlying potential. Consequently, understanding the behavior of soliton lattices can be approached through a generalization of the Frenkel-Kontorova (FK) model to quasiperiodic potentials \cite{van1999pinning,vanossi2000driven,braun2004frenkel}. Notably, we observe a commensurate-incommensurate transition at a critical field $H_{\rm CI}$. Below $H_{\rm CI}$, the number of solitons in the system increases in a semi-continuous manner, with a sequence of ground states differing by a single soliton. At zero field, the rotational symmetry around the chiral axis is restored, and the ground state manifests as a quasiperiodic helix.

\section{Summary and conclusions}
\label{sec:conclusions}

We have analyzed the ground state properties of a chiral helimagnet on a Fibonacci quasicrystalline lattice. We focused on an effective one-dimensional classical spin model with nearest-neighbor exchange and Dzyaloshinskii-Moriya interactions (DMI). An external magnetic field $H$, applied perpendicular to the helimagnet's axis, produces a diverse array of spin textures, depending upon the magnetic field intensity. These configurations include a non-collinear quasicrystalline quasi-fully polarized (QFP) state at high fields, and several chiral soliton lattices both commensurate and incommensurate with the underlying Fibonacci lattice. 

As the magnetic field decreases from the QFP phase, a critical field denoted as $H_c$ marks the onset of a Devil's staircase pattern characterized by chiral soliton quasicrystalline lattices commensurate with the underlying Fibonacci quasicrystal. A commensurate to incommensurate transition takes place at a lower field  $H_{\rm CI}$, giving rise to discommensurations within the otherwise quasicrystalline soliton lattice. With further reduction in magnetic field strength, the discommensurations proliferate, ultimately resulting in the formation of a quasiperiodic helical structure at zero field.

Over a broad spectrum of magnetic fields, encompassing the regime of quasiperiodic soliton lattices, the spin textures can be explained in terms of a dilute system (soliton size $l_s$ much smaller than average inter-soliton distance) of chiral solitons with  short-range  repulsive interactions and under the influence of an effective external quasiperiodic potential. This simple picture allows us to derive the intricate spin textures by treating the solitons as effective particles. Our methodology involves a systematic approach to constructing soliton lattices, which relies on two key steps: first, calculating the external potential acting on the solitons, and second, evaluating the interaction energy between solitons at various separations.
Once the positions of the solitons are determined, the orientations of the spins can be optimized accordingly. This method offers a notably more efficient strategy compared to the exhaustive minimization of spin orientations.

The effective potential governing the behavior of solitons gives rise to magnetic field plateaus characterized by a plethora of quasi-degenerate quasicrystalline soliton lattices. This extensive quasi-degeneracy originates from a quasiperiodic arrangement of neighboring pairs of soliton locations, each possessing similar energy levels.

While the numerical results presented are based on a specific parameter set, our key conclusions can be extended to a broader spectrum of parameters and related models. For instance, we can apply our findings to the $J_1$-$J_2$ model with easy-axis anisotropy, where the spiral states are not induced by the competition of nearest-neighbor ferromagnetic exchange $J_1<0$ with a Dzyaloshinskii-Moriya interaction, but with a second-neighbor  exchange  $J_2 > |J_1|/4$~\cite{cornaglia2023unveiling}. In  this  new scenario, the Hamiltonian does not determine the sign of vector chirality, implying  that solitons possess an internal  degree of freedom (vector chirality) that is also expected to organize in a certain manner. Particularly, we anticipate a comparable phase diagram where the intermediate field regime's behavior is governed by the interplay between an effective quasi-periodic potential for solitons and their short-range interactions. While varying parameters may alter the size of the solitons and the nature of the underlying pinning potential, we expect the qualitative features of the phase diagram as a function of external magnetic field to remain consistent across a wide parameter space.

In chiral helimagnets, itinerant electrons serve as a driving force for the formation and dynamics of chiral solitons~\cite{schulz2012emergent}. Conversely, the interaction between these magnetic structures and itinerant electrons generates an effective potential experienced by the electrons, which can be manipulated by external fields. This interaction suggests that magnetic structures in quasicrystals could offer a platform for studying the behavior of electrons in a quasiperiodic potential, allowing for tunable control over the system~\cite{Jagannathan21}.

In future investigations, we aim to expand this study to encompass two-dimensional quasiperiodic systems. Recently, two-dimensional quasicrystalline structures have been synthesized from 30$^\circ$ twisted bilayer graphene~\cite{ahn2018dirac}. We anticipate that magnetic van der Waals structures could serve as a platform for creating quasicrystalline magnetic systems. The insights gained from our results may contribute to the understanding of 2D topological defects within these systems.

\begin{acknowledgments}
P. S. C. acknowledges support from Grants PICT 2018-01546 and PICT 2019-00371 of the ANPCyT. This collaborative effort is made possible from generous support by Instituto Balseiro, Universidad de Cuyo, Fulbright Argentina, and through  a Fulbright Specialist award. The Fulbright Specialist Program, a part of the larger Fulbright Program, is a program of the Department of State Bureau of Educational and Cultural Affairs with funding provided by the U.S. Government and administered by World Learning. C. D. B. also acknowledges support from the U.S. Department of Energy, Office of Science, Office of Basic Energy Sciences, under Award Number DE-SC0022311.
\end{acknowledgments}

\appendix

\section{Energy minimization procedure} \label{app:minim}

To minimize energy and determine the ground state configuration for a given set of model parameters, we employ the Broyden-Fletcher-Goldfarb-Shanno (BFGS) method, as outlined in Ref.~\cite{cornaglia2023unveiling}, in conjunction with a genetic algorithm. The BFGS method is a quasi-Newton approach that utilizes an approximation of the Hessian matrix to locate the minimum of a function. While BFGS is generally reliable and efficient, it may become trapped in local minima, thereby failing to identify the global minimum. To mitigate this limitation, we augment the BFGS algorithm with a genetic algorithm.
We generate the genetic algorithm's initial population by running the BFGS method multiple times with randomly selected initial spin configurations. Subsequently, we select the configurations with the lowest energies to form the genetic algorithm's starting population.

The crossover between two spin configurations is performed by randomly selecting a range of spins in one of the configurations, and replacing it with the corresponding range of spins in the other configuration. Subsequently, the new configuration undergoes optimization using the BFGS algorithm. If the resulting configuration exhibits lower energy compared to the two original configurations, it is added to the population.
The procedure iterates until the population reaches a specified size. Subsequently, the configurations are sorted based on energy, and those with the highest energy are eliminated. This process continues until the population converges to a single configuration or reaches the maximum number of generations.

To validate the procedure, we conducted an independent run of the genetic algorithm using a different initial population to ensure consistency in obtaining the same ground state configuration.  For the intermediate range of external magnetic fields, up to $100,000$ initial BFGS runs were required.  The population size of the genetic algorithm was set to 1000, with a maximum of 100 generations allowed.

The bulk of our numerical calculations focused on chains consisting of 378 spins, utilizing open boundary conditions. However, to address finite size effects, we extended the minimization procedure to systems containing 611 and 988 spins as well. Remarkably, the magnetic field values corresponding to phase transitions between commensurate phases exhibited consistency across these diverse system sizes. This consistency can be attributed to several factors elucidated in the main body of the article. Specifically, the finite extent of single-soliton solutions, the short-range nature of soliton-soliton interactions, and the presence of an effective potential collectively enable the local determination of soliton configurations, irrespective of the system size.

\bibliography{solitons,magneticQC,assorted,commensurability-QC}
\end{document}